\documentclass[a4paper,12pt]{scrartcl}

\usepackage[utf8]{inputenc} 
\usepackage[T1]{fontenc}    
\usepackage{url}            
\usepackage{booktabs}       
\usepackage{amsfonts}       
\usepackage{nicefrac}       
\usepackage{microtype}      
\usepackage{lipsum}		
\usepackage{graphicx}
\usepackage{natbib}
\usepackage{doi}
\usepackage[a4paper,left=2cm,right=2cm,top=2.5cm,bottom=2.5cm]{geometry}

\hypersetup{
pdftitle={Intra-Chromosomal Potentials from Nucleosomal Positioning Data},
pdfauthor={Kunhe Li, Nestor Norio Oiwa,Sujeet Kumar Mishra, Dieter W. Heermann}
pdfkeywords={Polymer, chromosome, chromatin classification, nucleosome, effective potential, Monte Carlo method, C. albicans}
}

\begin{document}

\title{Intra-Chromosomal Potentials from Nucleosomal Positioning Data}


\author{Kunhe Li\\
Institute for Theoretical Physics, Heidelberg University \\
Philosophenweg 19, D-69120 Heidelberg, Germany\\
\\
Nestor Norio Oiwa\\
Department of Basic Science, Universidade Federal Fluminense \\
Rua Doutor S\'{i}lvio Henrique Braune 22, Centro, \\
28625-650 Nova Friburgo,  Brazil \\
\\
Sujeet Kumar Mishra\\
Center for Computational Biology and Bioinformatics\\
School of Computational and Integrative Sciences (SCIS) \\
Jawaharlal Nehru University \\
New Delhi, India\\
\\
Dieter W. Heermann\\
Institute for Theoretical Physics, Heidelberg University \\
Philosophenweg 19, D-69120 Heidelberg, Germany\\
\texttt{\small heermann@tphys.uni-heidelberg.de}
}

\maketitle

\begin{abstract}
No systematic method exists to derive intra-chromosomal potentials between nucleosomes along a chromosome consistently across
a given genome.  Such potentials can yield information on nucleosomal ordering, thermal as well as mechanical properties of 
chromosomes. Thus, indirectly, they shed light on a possible mechanical genomic code along a chromosome. To develop a method 
yielding effective intra-chromosomal potentials between nucleosomes a generalized 
Lennard-Jones potential for the parameterization is developed  based on nucleosomal positioning data. This approach eliminates some of the problems that
the underlying nucleosomal positioning data has, rendering the extraction difficult on the individual nucleosomal level. Furthermore, patterns on which to base a classification along a
chromosome appear on larger domains, such as hetero- and euchromatin. An intuitive selection strategy for the noisy-optimization problem 
is employed to derive effective exponents  for the generalized potential. The method is tested on the Candida albicans genome. Applying k-means clustering 
based on potential parameters and thermodynamic compressibilities, a genome-wide clustering of nucleosome 
sequences is obtained for Candida albicans.  This clustering shows that a chromosome beyond the classical dichotomic
categories of hetero- and euchromatin, is more feature-rich. 
\end{abstract}

\noindent Keywords

\noindent Polymer,  chromosome, chromatin classification, nucleosome, effective potential, Monte Carlo method, C. albicans

\section{Introduction}

The organization of a complex system such as the nucleosome
organization and with it the three-dimensional organization of a chromosome, 
is influenced by hundreds of factors from  DNA sequence,
nucleosome remodelers to transcription factors~\cite{Struhl:2013vh}. Each of these
factors influence not only the chemical environment but also, the mechanical
properties of the chromatin fiber such as the bending rigidity. Since the chromatin fiber
is a heteropolymer, the bending rigidity is not a constant along the backbone~\cite{Kumar:2018wr}. Changing the bending rigidity
by a more compact packing of the nucleosomes, for example by a micro phase separation~\cite{Conte:2020uh,Farr:2021wd} changing
the order parameter and packing, has an influence on the
loop structure of a chromosome and hence on regulation~\cite{Ghavi-Helm:2019wr}. 

It has long been speculated that there must be something like a mechanical code on top of the
genetic code~\cite{Zuiddam:2017tu,Basu:2020wi}. This mechanical code stems from the organizational
structure of the nucleosomes since  elasticity is a direct result of interatomic interaction.
A tighter packing gives rise to more steric repulsion and hence
higher bending rigidity. This in turn leads a reduced possibility for distal interactions, i.e., looping,
hence controlling the three-dimensional organizational structure. And, there is more and more 
evidence surfacing that there is a richer variety of compactification
of nucleosomes beyond the hetero- and euchromatin picture~\cite{Routh:2008aa,Bohn:2010vy,Tchasovnikarova:2018ue}. Experimental as 
well as theoretical work have indicated that indeed there is more than just two~\cite{Liu:2020wd,Hilbert:2021ve}.

In this work, we take the point of view that we can extract larger nucleosomal
structure from nucleosomal positioning data by coarse-graining.
To reveal the thermodynamic properties and hence give indication on the mechanical
code, we move to a larger global scale and ask for
nucleosomal distribution patterns along a single chromosome as well as universal
pattern between all chromosomes of a given genome. For this we need to eliminate some of the
smaller structures to reveal structure on a coarser level which is also more
in line with the local phase separation picture~\cite{Singh:2020ug}.

There are at least two main directions that can
be chosen. Physically, it is possible to start with geometric properties, e.g.,
the bending rigidity or stiffness, which is already verified to have a
significant correlation with the compaction~\cite{Segal2006,poirier2002bending}.
Chemically, it is desirable to extract the effective pair-wise potential between
single nucleosomes, and essential properties can be calculated subsequently.
This allows to compute thermodynamic properties such as the compressibility for
all of stretches showing a particular pattern of nucleosome distribution.
Eventually, this leads to information on the mechanical properties since it
allows to bring in line information on varying compressibilities and along the chromosomes with effective potentials. 
Furthermore, it also allows to extract the $\chi$-parameter for the Flory-Huggins theory and shed light on the possible
thermodynamic state, in particular the micro phase separation~\cite{Gibson:2019ua}.

\section{Methods}

\subsection{Computational Methods}

One of the basic techniques to measure the nucleosome activity is the
micrococcal nuclease digestion with deep sequencing
(MNase-seq)~\cite{klein2020genomic}. The method measures the nucleosome
occupancy by measuring the frequency of nucleosome bounded DNA
fragments. However, it does not directly identify the nucleosome
position, the probabilistic genomic position where each nucleosome is
located. In order to map the MNase-seq data to nucleosome positioning
data, several programs were developed, such as
NPS~\cite{schopflin2013modeling}, nucleR~\cite{flores2011nucler},
DANPOS~\cite{chen2013danpos}, and iNPS~\cite{chen2014improved} (improved nucleosome positioning from sequencing).

Our starting point is iNPS data for Candida albicans. The
raw data (MNase-seq) is available from the Gene Expression Omnibus (GSM1542419)~\cite{GEO:2021} 
and was measured by Puri et
al.~\cite{puri2014iron}. We also accessed the processed iNPS data
in the NucMap database by Zhao et al.~\cite{zhao2019nucmap}.

\begin{figure}[!ht]
   \centering
   \includegraphics[width=1.0\linewidth]{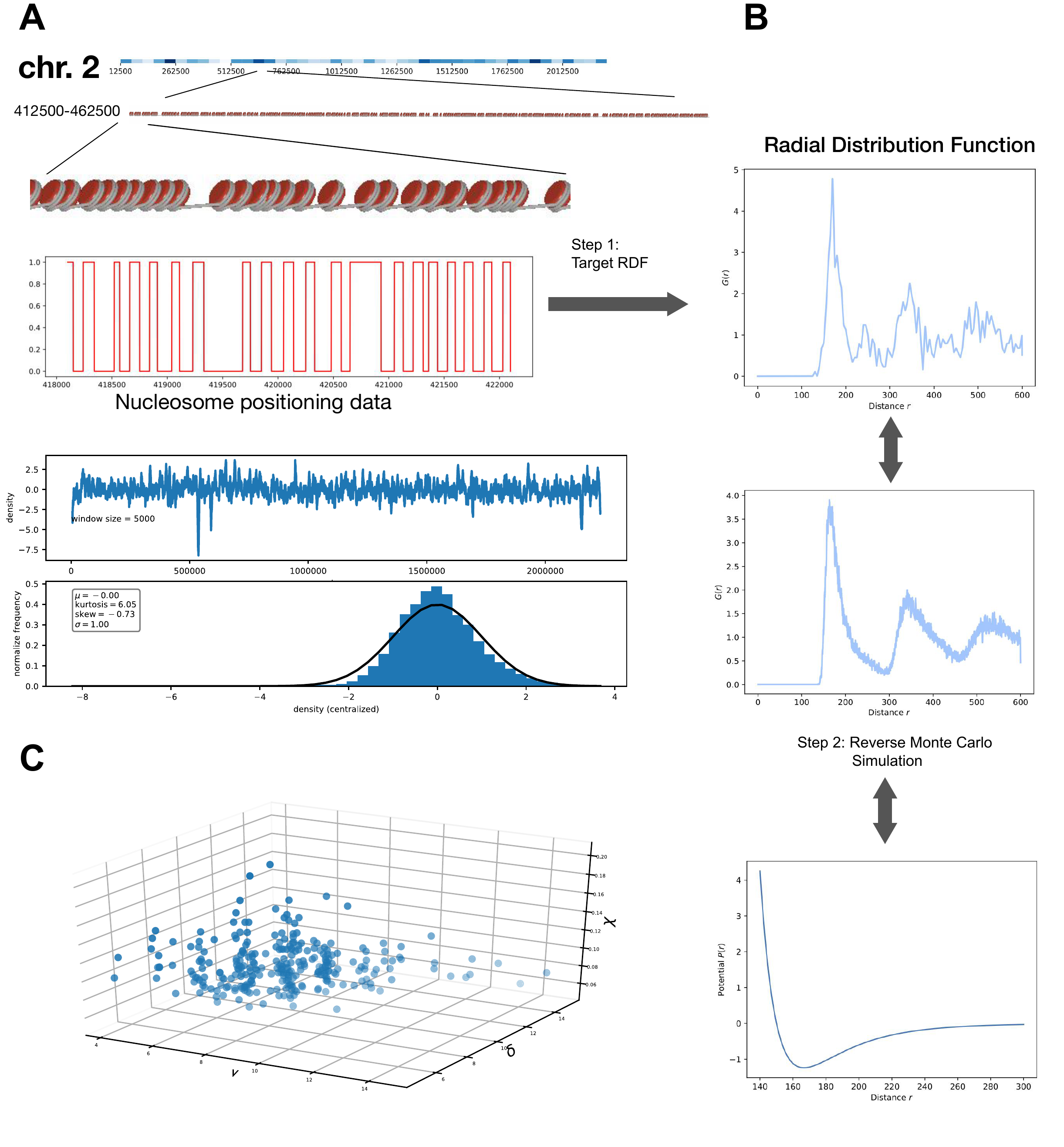}
\caption{\textbf{Steps to derive intra-chromosomal potentials from nucleosomal positioning data.}
Panel A shows schematically the distribution of nucleosomes in a section of
chromosome 2 of candida albicans.  We split the chromosome into sections,
typically of size $50000$ bp. The lower part of Panel A shows the density after applying a rolling mean
averaging with window size $5000$ bp, and the typical section size is chosen to be 10 times of this scale.
Step 1 takes the red binary data.
Based on this data, the radial distribution function (RDF) is computed. This step enables us to obtain a coarse-grained representation of the
chromosome that allows for an effective and efficient simulation of a
chromosome. There is also a $12500$ bp extra intersection at each end with its neighbor. This resolves the boundaries between the sections.
Once the radial distribution is computed, we apply a cut-off to the potential. Using a reverse Monte
Carlo simulation, we estimate a potential from the RDF.
We  employ an intuitive selection strategy, i.e., a noisy
optimization technique to find the best fit for the generalized Lennard-Jones
exponents (see Panel C).}
\label{Fig1}
\end{figure}

A section of the raw data is shown in Figure~\ref{Fig1} in  panel A
indicated by the red line. The areas with value $1$ are the nucleosome
positions, and the areas with value $0$ are voids. This data is noisy due to
missing data. Furthermore, on this small scale it is difficult to discern structure.

The goal is to extract potentials from the nucleosomal positioning data. One
approach to obtain those is to compute the radial distribution function (RDF)
$G(r)$ with respect to the distance $r$ (measured in base pairs)

\begin{equation}
\label{eq1}
G(r) = \frac{1}{\rho N S_d}\sum_{i=1}^{N}\sum_{j=1,j\neq i}^{N}\delta(r-r_{ij})
\end{equation}
where $\rho$ is the density, $N$ is the number of nucleosomes, $S_d$ is
a dimensional related term, $r_{ij}$ is the distance
between two nucleosomes $i$ and $j$ and $\delta(r-r_{ij})$ is equal to
$1$ if $r=r_{ij}$ and $0$ otherwise.

A chromosome is split into sections of $50000$ bp with $12500$ bp extra intersection at each end with its neighbour. For each section, we calculate the
corresponding RDF. The sectioning of the
chromosome is such that a substantial overlap between neighbouring sections is
guaranteed. Thus, the actual boundary position is somewhat fuzzy so that the actual
starting position becomes less relevant.

To derive pair-potentials from the nucleosomal distribution
patterns~\cite{heermann-2021} there are several paths such as the Berg-Harris
method~\cite{shimoji1967relation}, Yvon-Born-Green
equation~\cite{cho2009inversion} and reverse Monte
Carlo~\cite{lyubartsev1995calculation}. We  employ an reverse process on the nucleosomal radial distribution function. Its solution is guaranteed to
converge by combining the noisy
optimization~\cite{arnold2012noisy,mcnamara2000classification} with the
coarse-graining technique of molecular models, i.e., the reverse Monte
Carlo~\cite{lyubartsev2010systematic,Binder-Heermann:1988} and, for example,
implemented for the aqueous NaCl solution~\cite{lyubartsev1995calculation}. We
implemented the basic idea with several improvements: most importantly, a
generalized Lennard Jones model for the potential and an intuitive selection strategy (ISS) for the noisy optimization problem is used.

The reverse Monte Carlo (RMC) method is a double loop nested Monte Carlo (MC) simulation.
In the inner loop, a standard molecular Monte Carlo simulation is implemented to
obtain the desired parameter for a given potential. While for the outer loop 
a Monte Carlo Markov Chain (MCMC)~\cite{Binder2010} is employed. A MCMC step proposes a
new potential, runs the inner step, compares the computed parameter with
the target result, and updates the potential until the tolerance level is reached.
The RMC method succeeded in many cases, for example, in NaCl
solutions~\cite{lyubartsev1995calculation}. However, it has the flaw that it has
no guarantee to convergence, especially for a complex system. This issue also
emerged applying RMC for the nucleosome system. In this circumstance, we
have developed two improvements.

The original RMC uses a general potential. This, however,
leads to convergence problems. From the computed radial distribution
function $G(r)$ (Figure.~\ref{S2}) and the related mean-field potential

\begin{equation}
P_{\textrm{MF}}(r)\propto -\log(G(r)) \; ,
\end{equation}
we can actually observe that the target potential
has a type similar to a Lennard Jones potential.  Hence, without losing most of the
generality, our ansatz is a generalized Lennard Jones potential 

\begin{equation}
\label{eq2}
V(r) = 4\epsilon \left[ \left(\frac{\sigma}{r}\right)^\delta - \left(\frac{\sigma}{r}\right)^\nu \right] \; .
\end{equation}
Consistent with the Lennard Jones potential, $\epsilon$ determines the amplitude,
and $\sigma$ determines the length scale. The parameters $\delta$ and $\nu$ are the exponents,
that determine the shape and allows it to preserve most of the generality.

Another modification is substituting the MCMC step in RMC. The MCMC step is
intended to solve the optimization problem, i.e., finding the RDF minimizing the
differences. However, calculating an RDF from a potential via simulation
produces non-negligible noise, especially for a more complex system. Therefore,
the MCMC or other methods, e.g., Hill Climbing, Gradient Descent, and Simulated
Annealing, have low efficiency or are not converging. Consequently, we use for this
non-trivial step a noisy optimization technique  (dynamic
optimization~\cite{mcnamara2000classification}, or optimization with erroneous
oracles~\cite{singer2015information}). The straightforward application is via an
evolution strategy~\cite{arnold2012noisy}. We have modified this to an
intuitive selection strategy (ISS). This approach  is more stable and well-suited  for
parallel computing. Due to this parallelization, the computational cost is
strongly reduced. 

The ISS is very straightforward: 1. Execute the MC simulation for each possible
potential in low precision, i.e., smaller number of MC steps. 2. Choose the best
$N$ candidates according to a selection ratio $\theta$. 3. Increase the number of MC steps
to a larger value and repeat the process. Repeating this many times, finally,
there will be only one candidate, which is the result.

\subsection{Compressibility}

We compute the reduced isothermal compressibilities $\chi_T^\infty$ by the block
density distribution method~\cite{heidari2018fluctuations,rovere1988block}. In
this method, the whole section with size $L_0$ is separated into $M_b$ blocks.
The size of each block is $L=L_0/M_b$. Let $N$ be the number of
the nucleosomes in a block. If the distribution of $N$ is $P_{L,L_0}(N)$, its $k$-th moments $\left< N^k\right>_{L,L_0}$ is given by
\begin{equation}
\label{eq3}
\left< N^k\right>_{L,L_0}=\sum_N N^k P_{L,L_0}(N) \; .
\end{equation}
The summation is over all possible value of $N$. Then the reduced isothermal compressibility of a block is
\begin{equation}
\label{eq4}
\chi_T(L,L_0)=\frac{\left< N^2\right>_{L,L_0}-\left< N\right>_{L,L_0}^2}{\left< N\right>_{L,L_0}} \; .
\end{equation}
The difference between the finite size $\chi_T(L,L_0)$ and the thermodynamic limit $\chi_T^\infty$ is related to boundary 
effects associated with the finite-size of the subdomains. It takes the form:
\begin{equation}
\label{eq5}
\chi_T(L,L_0\to \infty)=\chi_T^\infty+\frac{c}{L}+O\left( \frac{1}{L^2}\right) \; .
\end{equation}
Here $c$ is a constant. Under this circumstance, the reduced isothermal compressibility of block $\chi_T(L, L_0)$ can 
be extrapolated to compute the reduced isothermal compressibility $\chi_T^\infty$ by just taking the limits $L, L_0\to \infty$.  
Hence, in the $\chi_T(L,L_0)$ vs. $M_b$ plot, the value at $M_b=0$ is the result $\chi_T^\infty$.

\subsection{Parameters}

For the each of the eight chromosomes of the genome we partitioned the chromosome in
sections of $50000$ bp length each.  There is a
$12500$ bp extra intersection at each end with its neighbor to reduce the
boundary effect. Thus, the total length of each section is $75000$ bp including the overlap. For the
particle-based Monte Carlo simulation section $i$ starts from $12500+50000\cdot i$ bp to
$12500+50000(i+1)$ bp, while actually the data is taken from $50000\cdot i$ bp
to $50000\cdot i+75000$ bp. This binning is applied to the whole genome. For
example, the length of chr. 2 is $2.231.883$ bp~\cite{rangwala2021accessing}, and
it is separated into $44$ sections.

In the one-dimensional Monte Carlo simulation, each monomer represents a nucleosome and occupies a volume equal to the averaged 
nucleosome length for that section. For every MC
step, a random move for each monomer is proposed. It ranges from 0 to $\lambda$.
The move is rejected or accepted according to the energy difference multiplied
by the Boltzmann factor $k_B T$. In our simulation, $k_B T$ is set
to be $1$. 

The value of $\lambda$ is chosen to be the smallest value
that allows the acceptance rate to be equal to or smaller than 50\% on average.

For the differences between the target RDF and the simulated we used  
the mean squared residual (MSR)

\begin{equation}
MSR = \frac{1}{(n-p)}\sum (x-\hat{x})^2 \; ,
\end{equation}
where $p$ is the number of parameters in the regression (including the intercept).
$x$ is the target value, and $\hat{x}$ is an estimator. 

For the modified Lennard-Jones potential the domain of $\sigma$ is $\left[140,170\right]$. It has the unit of
one base pair. Inside the ISS, the selecting ratio is 0.25. 

\subsection{Classification}

The resulting potentials  from the Monte Carlo with its parameters can  be used for clustering 
approaches such as k-means.  Panel C in Figure~\ref{Fig1} shows the obtained values for
the exponents as well as on the z-axis the compressibility data. The parameters $\nu$ and $\delta$
that characterize the short range repulsion and the long-range attraction together with the information
on the compressibility are used for a k-means clustering.

\section{Results}

\subsection*{Effective Potentials and Classification}

\begin{figure}[!ht]
   \centering
   \includegraphics[width=1.0\linewidth]{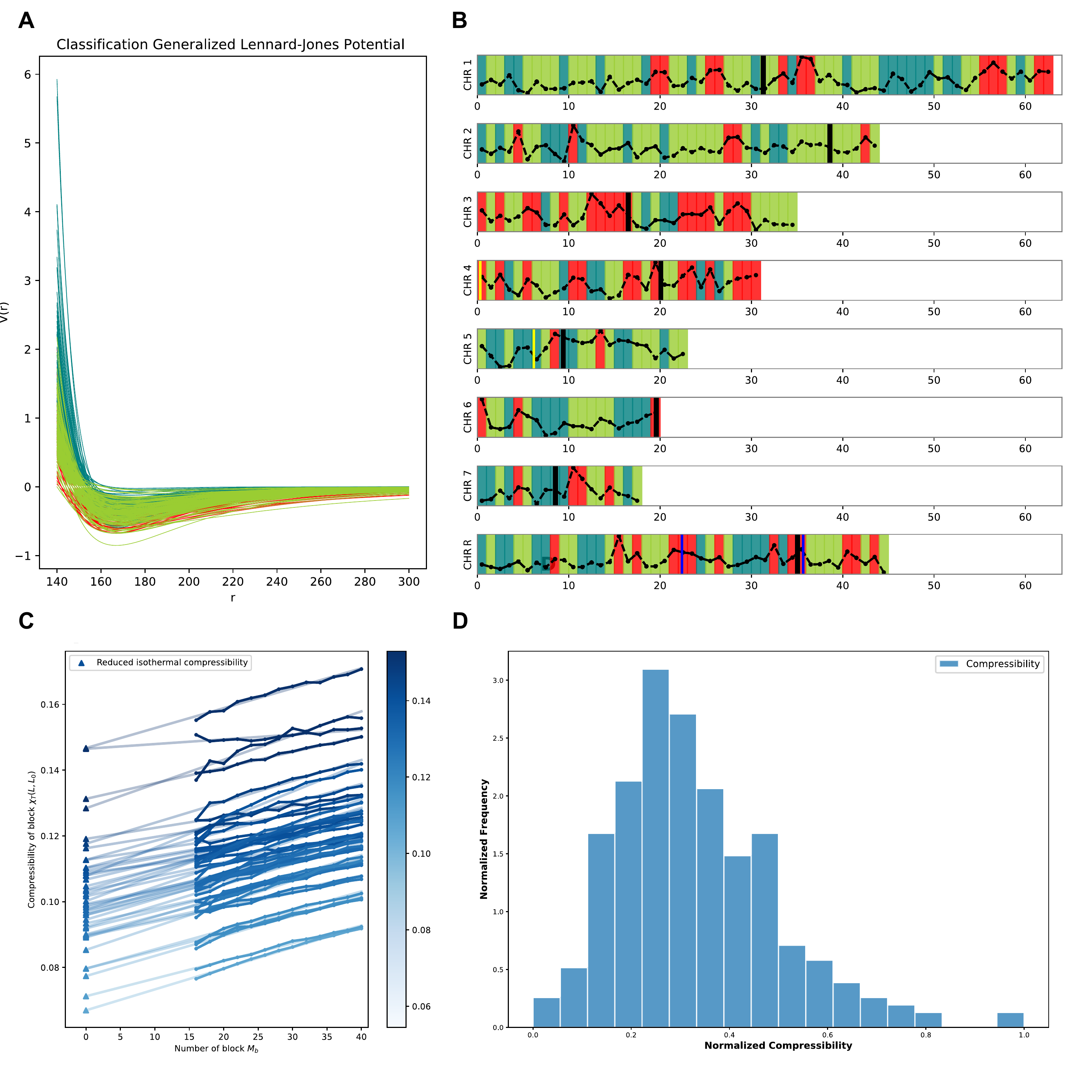}
\caption{\textbf{Effective pair-potential, genome-wide classification, and compressibility.}
\textbf{Panel A}: Shown is the result for Candida albicans. Each chromosome is partitioned into several
sections, each containing $50000$ base pairs with two additional $12500$ bp intersections on
both sides. The curves are the effective potentials, which quantify the global
interaction pattern between nucleosomes. 
Their coloring is adjusted to be consistent with panel B.
\textbf{Panel B} shows the classification of the sections based on the pair-potentials and compressibilities 
for the whole genome. This classification is based on a k-means clustering into $3$ 
clusters. They are intentionally classified to be comparable with the
classification of heterochromatin, euchromatin, and differently organized. 
The dashed lines are the compressibility results. The two yellow and the two
blue lines mark the position of known characterization.
\textbf{Panel  C}:
This panel shows the reduced isothermal compressibility $\chi_T^\infty$ employing
the block density method. The plot displays the process for chr. 2. The x-axis
is the number of blocks $M_b$. The linked dots are the compressibilities of
block $\chi_T(L, L_0)$. By extrapolating their linear regressions, we obtain the
intercepts as the compressibility, marked by triangles.  
\textbf{Panel D:} For a better representation of the
complex structure, we calculated the
distribution of the compressibility $P(\chi_T^\infty$).}
\label{Fig2}
\end{figure}

The results on the effective potential  for Candida albicans are shown in Figure~\ref{Fig2}(A). 
The colors indicate the class according to a k-means clustering based on three clusters taking
into account the exponents and the compressibility (see Figures~\ref{Fig1} and \ref{Fig2} Panels
C and D.)  From Figure~\ref{Fig2}(A), it can be seen that
they all share a minimum lying between $160$ bp to $180$ bp. However, the well depths are 
falling into different classes. A shallow minimum with a steep repulsive part indicates an area where 
nucleosomes  are loosely bound, corresponding to an irregular array, i.e. with liquid-like structure. 
A deep minimum with a 
less steep repulsion leads to a regular array in contrast, i.e. a much more ordered structure. Thus, the section partition into 
those that are liquid- and those that are more solid-like in agreement with the classical classification
hetero- and eu-chromatin picture, disregarding the nuances of a finer partitioning. However, the
classification did no trivially sort the potentials according to the potential minima. Rather, an interplay
between attraction, repulsion and compressibility can be seen. The sorting into classes is more towards
how the potential behaves at short distances and a larger distances. Whereas in the well part of
the potential a substantial criss-crossing can be seen the far ends are much more sorted.

The classification is based on all of the sections of the entire genome. This effectively constraints the pattern 
to be of a universal genome-wide character. Local variations are subsumed into broader classes 
filtering out the universal patterns underlying the
local variations within a chromosome as well as among the chromosomes.   

The resulting coloring of three clusters is shown in Figure~\ref{Fig2}(B). The coloring of 
Figure~\ref{Fig2}(A) is adjusted to be consistent with that in panel B.  The classification results
suggests that there is more than hetero- and eu-chromatin. At least a further class can be 
distinguished genome-wide. In the supplementary information Figure~\ref{S3} shows a principal
component analysis for various given k-means clusterings. Since we cannot employ directly
a method such as the elbow method to look for the best classification, the visual inspection
partitioning of the clusters in principal component space is used.  A classification into three clusters
shows the best result. Two clusters shows a trivial partition while for a larger number of 
clusters a significant overlap is seen. Indeed, already in the first experiments it was noticed
that within hetero- and euchromatin variations exist~\cite{Allshire:2018va}. 

The result of the classification into three classes mapped to their original genomic location
is shown in panel B of Figure~\ref{Fig2}. Also shown in the figure are the results for the
compressibility. The compressibilities themselves are shown in panel C and D. 
In Figure~\ref{Fig2}(C), we show the results from the block density method for all sections in chr. 2. 
Each line presents one section. The linked dots are the reduced isothermal
compressibility of block $\chi_T(L, L_0)$ with respect to the number of blocks $M_b$.
The straight lines are the corresponding linear regression results for the extrapolation 
to the thermodynamic limit. The triangles mark the intercepts, i.e., the reduced isothermal 
compressibilities $\chi_T^\infty$. All lines are colored according to their $\chi_T^\infty$ value.
Note that no corrections for the scaling are necessary as the extrapolation proportional
to $1/L$ is consistent with the data.

The distribution of the extrapolated compressibility values  for the whole genome 
(for Candida albicans) is shown in Figure~\ref{Fig2}(D). The distribution is clearly non-gaussian.
The obtained extrapolated values are used for the classification and shown in panel B. A high value of compressibility 
is associated with a few location along the chromosomes. Marked by the thick black line is the 
location of the centromeres. Four further markers from gene expression results confirmed by three experimental groups\cite{puri2014iron,chen2011iron,lan2004regulatory} are also included. 
They have measured their response to iron and concluded that the two blue marked regions were 
suppressed while the yellow marked regions were not suppressed. Both results are compatible with
the classification. The sections that are classified as heterochromatin are indeed consistent with the deeper wells of the potentials while the euchromatic region are in general 
associated with more shallow wells of the potentials.

\section{Conclusion}

Based on the nucleosomal positioning data the extraction of effective potentials is possible for an entire
genome. If this information is supplemented with thermodynamic information in terms of compressibility,
i.e., density fluctuations a genome-wide consistent classification in sections is possible. The classification
into the classes shows that at least three different classes must exist. Hence beyond hetero- and eu-chromatin
a third kind of ordering is necessary. The grouping of the exponents of the generalized Lennard-Jones
potential may suggest that there may be more than three classes. However, the principal component
analysis of the parameters into two dimensions shows that at least for this projection three is the
best decomposition into classes. 

Positioning data and simulations of the fluctuations of the positioning data should incorporate such effects
as nucleation of hetero-chromatic regions. Thus, in a consistent manner the classification into more or less
ordered regions is possible. Beyond this classification, having the information on the coarse-grained
potentials, this approach allows for the modelling of chromosomes as hetero-polymers with intera-chromosonal 
interactions. If this is further augmented with inter-chromomal information derived from chromosomal
conformation capture methods, a consistent framework for the simulation of chromosomes with the
effective potentials is possible. This then allows to look for the mechanics, i.e., the mechanical code.
Having the information on the potentials enables the modeling of the nucleosomes
as effective disks such that the steric interactions together with the density fluctuations yields information
on the stiffness of the particular section and thus on its bending rigidity. 

One aspect of the ordering and stiffness of segments that is not yet covered by the approach
are methylation effects. However, this can in principal be incorporated if a consistent set of
experimental data would be available for a particular genome. This would add a further dimension for the 
classification.

\medskip
\textbf{Acknowledgements} \par 
This work is funded by the Deutsche Forschungsgemeinschaft (DFG, German Research Foundation) under Germany's Excellence Strategy EXC 2181/1 - 390900948 (the Heidelberg STRUCTURES Excellence Cluster).
The authors gratefully acknowledge the data storage service SDS@hd supported by the Ministry of Science, Research and the Arts Baden-W\"urttemberg (MWK) and the German Research Foundation (DFG) through grant INST 35/1314-1 FUGG and INST 35/1503-1 FUGG. Kunhe Li would  to acknowledge funding by the Chinese Scholarship Council (CSC). Sujeet Kumar Mishra would like
to acknowledge funding by the India government Department of Biotechnology (DBT)-Interdisciplinary Research Center for Scientific Computing (IWR) PhD program.

\medskip

%

\bibliographystyle{unsrt}
\bibliography{bib}

\clearpage
\section*{Supplemental Information}
The source code for the program is available at https://github.com/mdscolour/reverseMC. 

The genome-wide effective potential data as well as the corresponding compressibility is available at the following DOI link: https://doi.org/10.11588/data/H3KPEU.

\renewcommand{\thefigure}{S\arabic{figure}}
\renewcommand{\thetable}{S\arabic{table}}
\setcounter{figure}{0}

\begin{figure*}[!ht]
   \centering
   \includegraphics[width=0.45\linewidth]{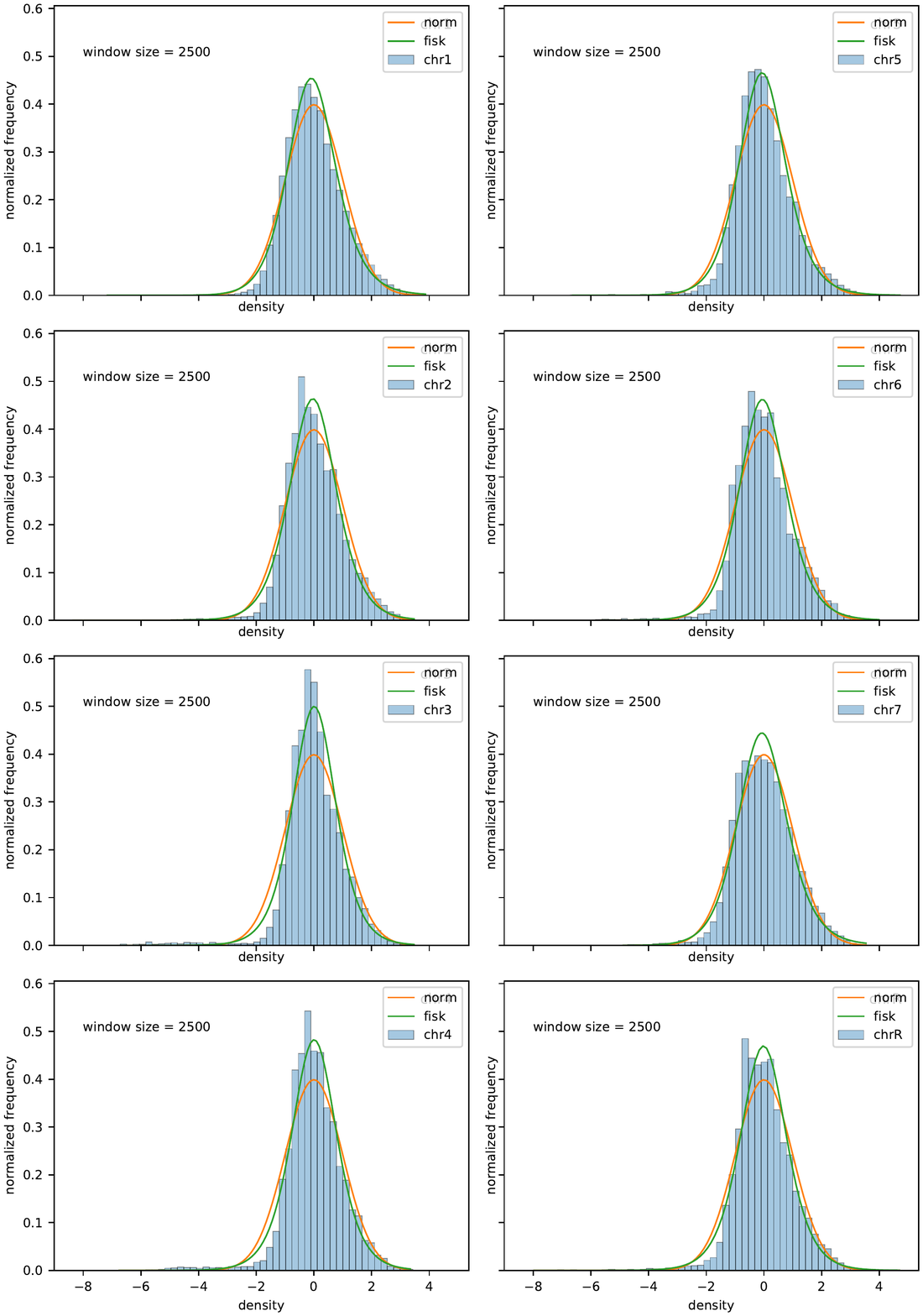}
   \includegraphics[width=0.45\linewidth]{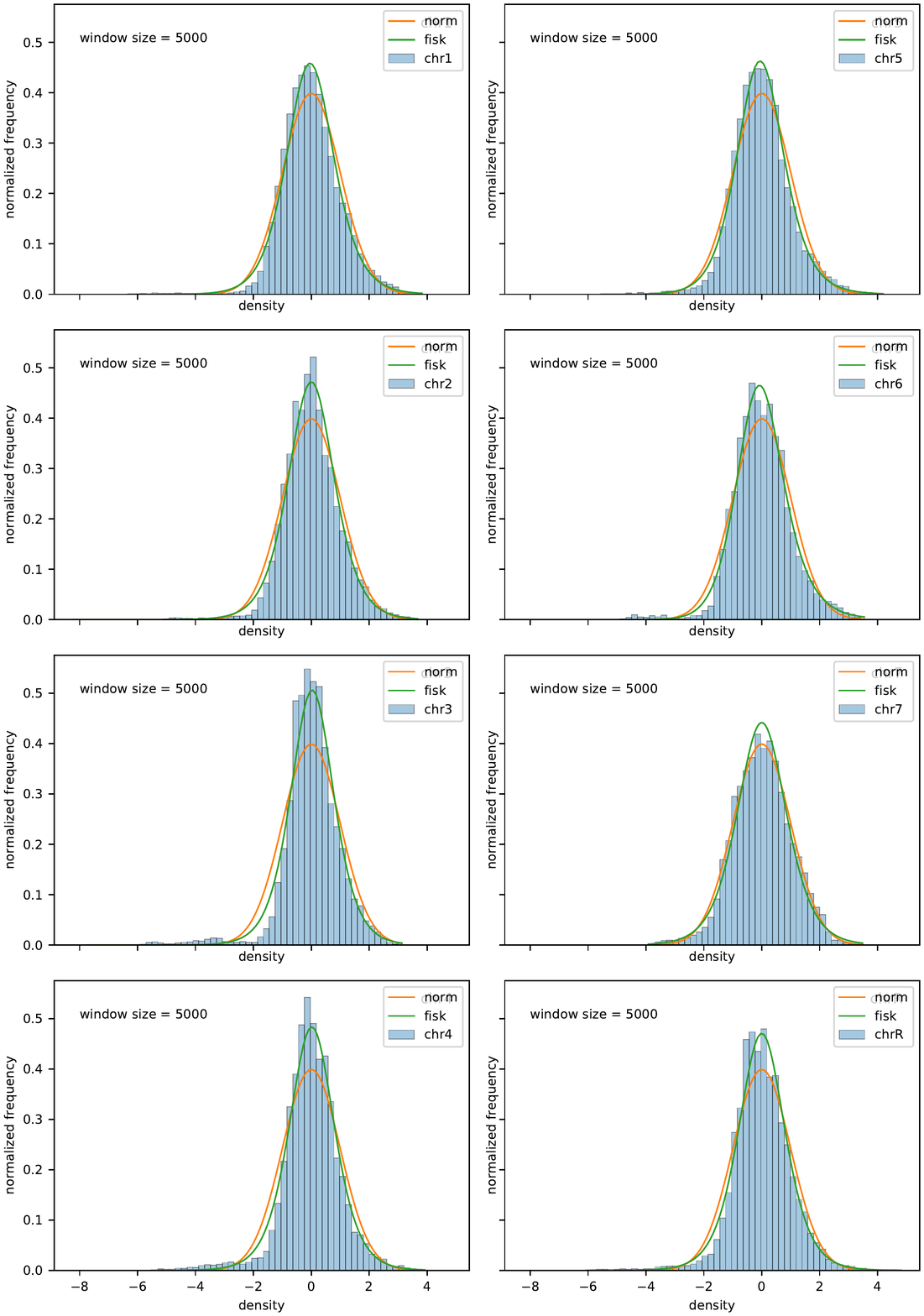}
\caption{ \textbf{Nucleosomal density after coarse-graining}
The nucleosomal density distribution appears to be close to a Fisk distribution, i.e.  is 
a log-logistic distribution. Shown are the results for a window size of $2500$ and $5000$. Several window sizes are examined and the $5000$ bp length is the most suitable coarse-graining scale. Hence the typical section length is chosen to be $50000$ bp.}
\label{S1}
\end{figure*}

\begin{figure*}[!ht]
   \centering
   \includegraphics[width=1.0\linewidth]{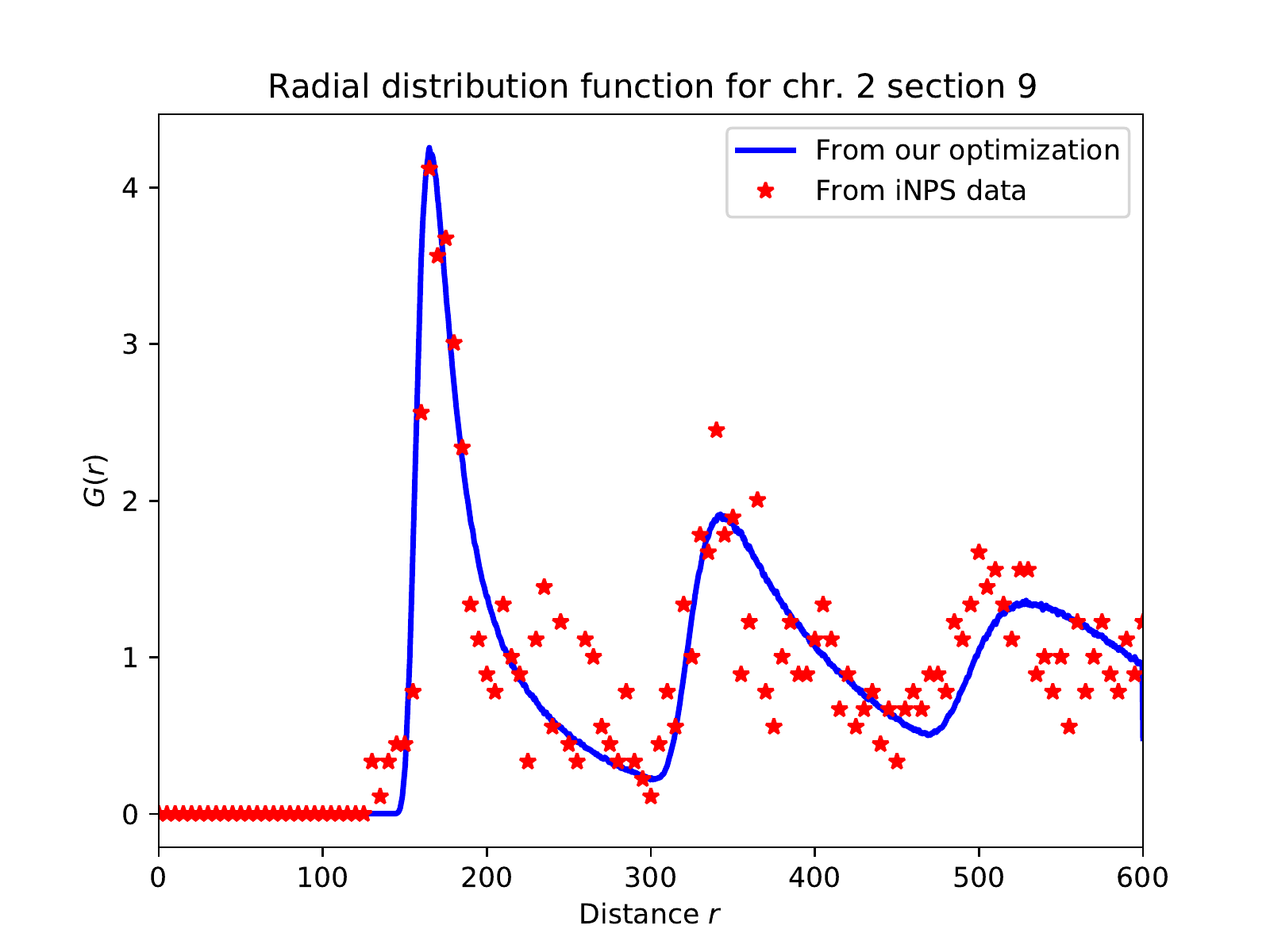}
\caption{\textbf{iNPS data and resulting radial distribution function for chr. 2 section 9}
Red stars show the radial distribution function (RDF) data calculated from
experimental iNPS data. The blue curve is the estimated result for the
effective potential at the same area by implementing An MC simulation. The RDF is computed from a total of $150000$ MC steps. 
}
\label{S2}
\end{figure*}

\begin{figure*}[!ht]
   \centering
   \includegraphics[width=1.0\linewidth]{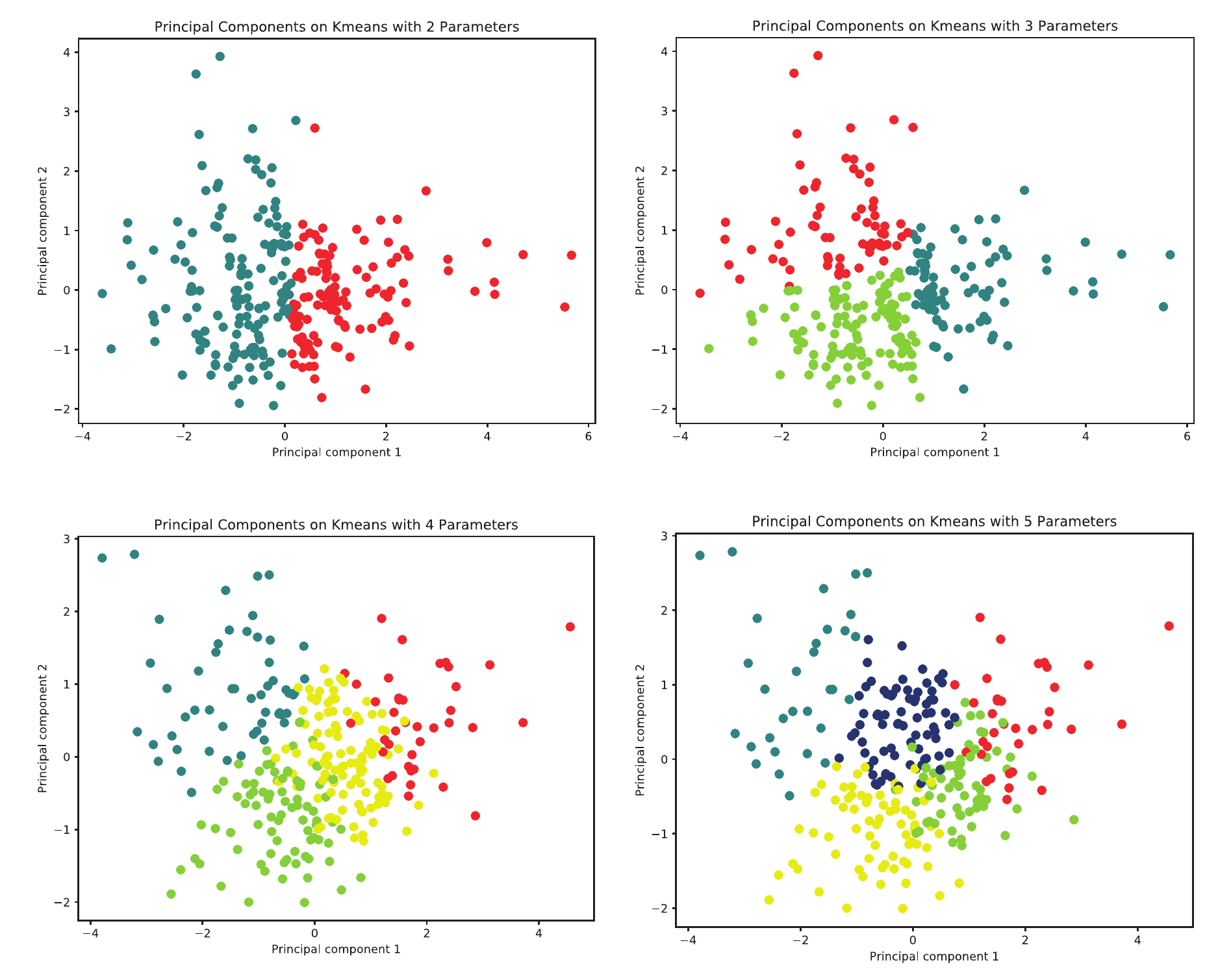}
\caption{\textbf{PCA on K-means}
Shown are the principal component analysis for a two-dimensional projection of the k-means clustering 
with various given cluster numbers. Rather than using the elbow or similar methods to find the optimum number
of clusters, we have chosen to visually detect the best number of clusters. From the visual inspection
we see that two clusters trivially separate into two clusters. The cluster separate non-trivially for three clusters whereas,
above three clusters there is always a non-negligible overlap between the clusters. The parameters for
the k-means clustering were $\nu$, the potential minimum and the compressibility $\chi$ within the section.
}
\label{S3}
\end{figure*}

\end{document}